\documentclass{article}[11pt]
\usepackage{a4wide}
\usepackage{bm}
\usepackage{bbm}
\usepackage{epsfig,graphics}
\usepackage{amsmath,amssymb,amsfonts}
\usepackage{color}
\usepackage{epstopdf}
\usepackage[
colorlinks=true,
linkcolor=black,
breaklinks=true,
urlcolor=green,
citecolor=blue]{hyperref}

\usepackage{wrapfig}
\usepackage{amssymb,amsfonts,bm}
\usepackage{dsfont}

\newcommand{\be}{\begin{equation}}
\newcommand{\ee}{\end{equation}}
\newcommand{\ba}{\begin{eqnarray}}
\newcommand{\ea}{\end{eqnarray}}

\renewcommand{\Im}{\textrm{Im}\,}

\title{\bf Two-pole structures in QCD: Facts, not fantasy!}

\author{Ulf-G. Mei{\ss}ner\\[0.3em]
{\em Helmholtz-Institut f\"ur Strahlen- und Kernphysik}\\ {\em and Bethe
Center for Theoretical Physics,}\\
{\em Universit\"at Bonn, D-53115 Bonn, Germany}\\
\and
{\em Institute for Advanced Simulation, Institut f\"ur Kernphysik,}\\
{\em and J\"ulich Center for Hadron Physics,}\\ {\em Forschungszentrum J\"ulich,
D-52425 J\"ulich, Germany}\\
\and
{\em Tbilisi State University, 0186 Tbilisi, Georgia}}

\begin{document}


\maketitle

\begin{abstract}
The two-pole structure refers to the fact that particular single
states in the spectrum as listed in the PDG tables are often  two states.
The story began with the $\Lambda(1405)$, when in 2001, using unitarized chiral perturbation
theory, it was observed that there are two poles in the complex plane, one close to
the $\bar{K}p$ and the other close to the $\pi\Sigma$ threshold. This was later understood
combining the SU(3) limit and group-theoretical arguments.  Different unitarization approaches
that all lead to the two-pole structure have been considered in the
mean time, showing some spread in the pole positions. This fact is now part of the PDG book,
though it is not yet listed in the summary tables.
Here, I will discuss the open ends and critically review approaches that
can not deal with this issue. In the meson sector some excited
charm mesons are good candidates for such a two-pole structure. 
Next, I consider in detail the $D_0^*(2300)$, that is another candidate
for this scenario. Combining lattice QCD with chiral unitary approaches
in the finite volume, the precise data of the Hadron Spectrum
Collaboration for coupled-channel $D\pi$, $D\eta$, $D_s\bar{K}$
scattering in the isospin $I=1/2$ channel indeed reveal its two-pole
structure. Further states in the heavy meson sector with
$I=1/2$ exhibiting this phenomenon are predicted, especially in the beauty
meson sector. I also discuss the relation of these two-pole structures and the
possible molecular nature of the states under consideration.
\end{abstract}

\section{Introduction}

The hadron spectrum is arguably the least understood part of Quantum Chromodynamics (QCD), the
theory of the strong interactions. It is part of the successful Standard Model (SM) and thus, we
can say that  structure formation, that is the emergence of hadrons and nuclei from the
underlying quark and gluon degrees of freedom, is indeed the last corner of the SM that is not yet
understood. For a long time, the quark model of Gell-Mann~\cite{GellMann:1964nj}
and Zweig~\cite{Zweig:1981pd} (and many sophisticated extensions thereof, such as \cite{Godfrey:1985xj}) 
have been used to bring order into the particle zoo. However, already before
QCD it was a puzzling fact that all observed hadrons could be described by the simplest combinations
of quarks/antiquarks, namely mesons as quark-antiquark states and baryons as three quark states,  while symmetries and quantum numbers would also
allow for tetraquarks, pentaquarks and so on. The situation got even worse when QCD finally appeared on
the scene, as it allows for the following structures (bound systems of quarks and/or gluons):
\begin{itemize}
\item Conventional hadrons, that is mesons and baryons as described before;
\item Multiquark hadrons, like tetraquark states (mesons from two quarks and two antiquarks),
  pentaquark states (baryons made from four quarks and one antiquark), and so on;
\item Hadronic molecules and atomic nuclei, that is multiquark states composed of a certain
  number of conventional hadrons (as discussed in more detail below);
\item Hybrid states, which are composed of quarks and (valence) gluons;
\item Glueballs, bound states solely made of gluons, arguably the most exotic form of matter,
  which has so far been elusive in all searches.
\end{itemize}  

The observed hadrons are listed with their properties in the tables of the Particle Data
Group (PDG) (also called ``Review of Particle Physics'' (RPP))~\cite{Tanabashi:2018oca}
within a certain rating scheme, just telling us that some states are better understood as others.
What complicates matters a lot is the fact that almost all hadrons are {\bf resonances}, that is unstable states.
These decay into other hadrons and leptons, like e.g. the $\rho$ meson decays into two pions or into a pion
and a photon (and other final states). Such a resonance is thus described by a complex energy, more
precisely, the real part is called the mass, $m_R$, and the imaginary part the half-width, $\Gamma_R/2$.
Consequently, all other properties are also given by complex numbers.
The only model-independent way\footnote{There are very few exceptions of
  isolated resonances on an energy-independent background where other methods can be used, but
  even in such cases a unique determination of the mass and the width is not always possible,
  see e.g. the discussion of the $\rho(770)$ in the RPP.}
to pin down these basic resonance properties is to look for poles in the complex plane, where
resonances are usually located on the second Riemann sheet at
\begin{equation}
z_{R} = (\Re z_R, \Im z_R) =  (m_{R}, \Gamma_R/2)~.
\end{equation}  
The residues at these poles contain information about the possible decays of such a resonance,
as will be discussed in more detail below.

As I will discuss in what follows, the hunt for such poles in the complex plane has revealed the
astonishing feature of the {\bf two-pole structure}, namely that certain states that are listed in the
RPP are indeed superpositions of two states. The most prominent example is the $\Lambda(1405)$,
which is discussed in detail in Sect.~\ref{sec:L1405}. More recently, the $D_0^*(2300)$, an excited charm meson,
has become another prime candidate for the two-pole scenario, paving the way for a whole set of such
states in the heavy-light sector (mesons made of one light $(u,d,s)$ and one heavy $(c,b)$ quark), see
Sect.~\ref{sec:D2300}. Before discussing these intricate states, it is, however, necessary to review
the pertinent methods underlying the theoretical analyses, see Sect.~\ref{sec:meth}. The conclusions and outlook
are given in Sect.~\ref{sec:summ}.

\section{Methods}
\label{sec:meth}

We start with the Lagrangian of QCD for up, down and strange quarks ($N_f=3$), which can be written as:
\begin{eqnarray}
{\cal L}_{\rm QCD} &=&  {\cal L}^0_{\rm QCD} - \bar{q} {\cal M} q~, \nonumber\\ 
{\cal L}^0_{\rm QCD} &=& -\frac{1}{2g^2}{\rm Tr}\left[ G_{\mu\nu}G^{\mu\nu}\right] + \bar{q} i \gamma^\mu
(\underbrace{\partial_\mu - i A_\mu}_{= D_\mu}) q~.  
\end{eqnarray}  
Here, $q=(u,d,s)^T$ is the quark triplet, $A_\mu$ is the gluon field, $G_{\mu\nu}$ the 
gluon field strength tensor, $g$ is the strong coupling constant and the color indices related to the
underlying SU(3)$_c$ local gauge symmetry have not been displayed. Further, ${\cal M} = {\rm diag}(m_u,m_d, m_s)$
is the quark matrix and the heavy flavors charm and bottom can be added analogously\footnote{We eschew here
  the top quark as it does not form hadrons due to its fast decay.}. Also, gauge fixing and the CP-violating
$\theta$-term are not displayed.  Remarkably, ${\cal L}^0_{\rm QCD}$ displays a SU(3)$\times$SU(3) flavor
symmetry (I do not discuss the additional U(1) symmetries/non-symmetries here),
\begin{equation}
{\cal L}^0_{\rm QCD} (G_{\mu\nu}, q', D_\mu q') = {\cal L}^0_{\rm QCD} (G_{\mu\nu}, q, D_\mu q)~,
\end{equation}   
in terms of left- and right-handed quark fields,
\begin{equation}
  q' = g_R P_R q + g_LP_Lq~,~~P_{R,L}=\frac{1}{2}(1\pm \gamma_5)~,~~g^{}_Ig^\dagger_I = \mathds{1}~,
  ~~{\rm det}g_I = 1~,~~I = L,R~.
\end{equation}
This is the {\bf chiral symmetry} of QCD. It leads to $16 = 2\cdot(N_f^2-1)$ conserved Noether currents,
that can be rearranged as 8 conserved vector and 8 conserved axial-vector currents. However, we know that the
symmetry of the groundstate is not the symmetry of the QCD Hamiltonian, as e.g. there is no parity-doubling
in the spectrum. The symmetry is spontaneously broken (or hidden, as Nambu preferred to say) to its vectorial
subgroup:
\begin{equation}
  {\rm SU(3)}_L \times {\rm SU(3)}_R \to {\rm SU(3)}_V~.
\end{equation}
The Goldstone theorem then tells us that for each broken generator there should be
a massless boson (the famous Goldstone bosons). Therefore, in the absence of quark masses,
we are dealing with a theory without a mass gap, which implies that Taylor expansions are
not analytic. When the quark masses are included, the pseudoscalar Goldstone bosons acquire
a small mass. In fact, the lightest hadrons are the eight pseudoscalar mesons ($\pi, K, \eta$).
All this is the basis for the formulation of an effective field theory (EFT), that allows for
perturbative calculations at low energy. Similarly, for the heavy quarks $c$ and $b$ one can
formulate a different EFT, based on the fact that the $c$ ad $b$ masses are large, $m_{c,b} \gg \Lambda_{\rm QCD}$,
as discussed next.

\subsection{Limits of QCD}

We have just discussed one particular limit of QCD, namely the chiral limit of the three light flavor
theory. Such a special formulation can be extended also to the heavy quark sector and to so-called
heavy-light systems. The various limits of QCD are:
\begin{itemize}
\item {\bf Light quarks:}
\begin{equation}
{\cal L}_{\rm QCD} = \bar q_L \, i D\!\!\!\!/ \, q_L 
+ \bar q_R \, i D\!\!\!\!/ \, q_R 
+ {\cal O}(m_f / \Lambda_{\rm QCD})~~~[f=u,d,s]~.
\end{equation}
In this limit, left- and right-handed quarks decouple which is the  chiral symmetry. As stated, it is spontaneously
broken leading to the appearance of 8 pseudo-Goldstone bosons. The pertinent EFT is chiral perturbation
theory (CHPT), see Sect.~\ref{sec:chpt}. Note that the  corrections due to the quark masses are powers in $m_f$.
\item {\bf Heavy quarks:}    
  \begin{equation}
   {\cal L}_{\rm QCD} = \bar Q_f \,  i v \cdot D \, Q_f  + {\cal O}(\Lambda_{\rm QCD} / m_f)~~~[f=c,b]~,
  \end{equation}
  where $Q$ denotes the field of a heavy quark.
  In this limit, the Lagrangian is independent of quark spin and flavor, which leads to  SU(2) spin and SU(2)
  flavor symmetries, called HQSS and HQFS, respectively. The pertinent EFT is  heavy quark effective field theory
  (HQEFT), see e.g.~\cite{Neubert:1993mb,Manohar:2000dt}. Here, the corrections due to the quark masses
  are powers in $1/m_Q$.
\item {\bf Heavy-light systems:}~Here,  heavy quarks act as matter fields coupled to light pions and one
  thus can combine CHPT and HQEFT as pioneered in~\cite{Burdman:1992gh,Wise:1992hn,Yan:1992gz}, see also
  Sect.~\ref{sec:HL}.
\end{itemize}  

\subsection{A factsheet on chiral perturbation theory}
\label{sec:chpt}
Chiral perturbation theory is the EFT of QCD at low energies~\cite{Weinberg:1978kz,Gasser:1983yg}.
For introduction and reviews, see e.g.~\cite{Ecker:1994gg,Pich:1995bw,Bernard:2006gx}.
Its basic properties are:
\begin{itemize}
\item ${\cal L}$ is symmetric under some Lie group ${\cal G}$, here ${\cal G}$ = SU(3)$_L \times$ SU(3)$_R$.
\item The ground state $|0\rangle$ is asymmetric and ${\cal G}$ is spontaneously broken to ${\cal H} \subset {\cal G}$,
  leading to the the appearance of Goldstone bosons (GBs) $|\phi^i(p)\rangle$. In QCD,  ${\cal H}$ = SU(3)$_V$
  and the Goldstone bosons are the aforementioned eight pseudoscalar mesons.
\item  In QCD, the matrix element of the axial-vector current ${\cal A}_\mu^i$,
  $\langle 0|{\cal A}_\mu^i|\phi^k(p)\rangle = i \delta^{ik}p_\mu F \neq 0$ $(i=1,\ldots,8)$, where $F$ is related
  to the pseudoscalar decay constant in the chiral limit.  $F\neq 0$ is a sufficient and necessary condition
  for spontaneous chiral symmetry breaking.
\item There are no other massless strongly interacting particles.  
\end{itemize}
Universality tells us that at low energies, any theory with these properties looks the same as
long as the number of space-time dimensions is larger than two. One can readily deduce that the
interactions of the GBs are weak in the low-energy regime and indeed vanish at zero energy. This
allows for a systematic expansion in small momenta and energies, and the quark masses lead to
finite but small GB masses, which defines a second expansion parameter. In fact, these two parameters
can be merged in one. The corresponding effective Lagrangian is readily constructed, it takes the form
\begin{equation}
{\cal L}_{\rm eff} = {\cal L}^{(2)} + {\cal L}^{(4)} + {\cal L}^{(6)} + \ldots \, ,
\end{equation}
where the superscript denotes the power of the small expansion parameter $p$ (derivatives and/or GB
mass insertions). This expansion is systematic, as an underlying power counting~\cite{Weinberg:1978kz} can be
derived. This shows that graphs with $n$ loops are suppressed by powers of $p^{2n}$ and that
at each order, we have local operators accompanied by unknown coupling constants, also called
low-energy constants (LECs). These LECs must be determined from fits to experimental data or
can be calculated using lattice QCD. Their specific values single out QCD from the whole universality
class of theories discussed above. One important issue concerns unitarity. Leading order calculations are
based on tree diagrams with insertions from ${\cal L}^{(2)}$, which means that such amplitudes are real.
Imaginary parts are only generated at one-loop order through the loop diagrams, which means that
unitarity is fulfilled perturbatively but not exactly in CHPT, for a general discussion, see~\cite{Gasser:1990bv}.
We will come to this issue in Sect.~\ref{sec:uni}.

Matter fields like baryons can also be included in a systematic fashion. There is one major complication,
namely the matter field mass that is of the same size as the breakdown scale of the EFT, here $\Lambda_\chi
\sim 1\,$GeV. Therefore, only three-momenta of the matter fields can be small and the mass must be dealt
with in some manner. Various schemes like the heavy baryon approach~\cite{Jenkins:1990jv,Bernard:1992qa},
infrared regularization~\cite{Becher:1999he} and the extended on-mass-shell scheme~\cite{Fuchs:2003qc}
exist to restore the power counting. For details, I refer to the
reviews~\cite{Bernard:1995dp,Bernard:2007zu,Scherer:2012zzc}.

\subsection{Chiral perturbation theory for heavy-light systems}
\label{sec:HL}

In this section, we display the effective Lagrangians that we need for the discussion of charm mesons
and their interactions. Consider first Goldstone boson scattering off $D$-mesons. The effective Lagrangian
takes the form~\cite{Cheng:1993kp,Lutz:2007sk,Guo:2008gp}:
\begin{eqnarray}
\label{LDphi}  
{\cal L}_{\rm eff} &=& {\cal L}^{(1)} + {\cal L}^{(2)}~,\nonumber\\
{\cal L}^{(1)} &=& {\cal D}_\mu D {\cal D}^\mu D^\dagger - M_D^2 DD^\dagger~,\nonumber\\ 
{\cal L}^{(2)} &=& D\left[-h_0 \langle\chi_+ \rangle
 -h_1 \chi_+ + h_2 \langle u_\mu u^\mu\rangle -h_3 u_\mu u^\mu\right] \bar{D}
  + {\cal D}_\mu D \left[ h_4 \langle u^\mu u^\nu\rangle - h_5 \{u^\mu,u^\nu\}\right]{\cal D}_\nu \bar{D}~.
\end{eqnarray}
Here, $D = (D^0,D^+,D_s^+)$ is the $D$-meson triplet, $M_D$ the average mass of the $D$-mesons,
and we utilize the standard chiral building blocks $u_\mu \sim \partial_\mu \phi$, with $\phi$ a member of
the GB octet, and $\chi_+ \sim {\rm diag}(m_u,m_d,m_s)$. The pertinent LECs can all be determined:
$h_1 = 0.42$ from the $D_s$-$D$ splitting, while $h_{2,3,4,5}$ are fixed from a fit to lattice
data~\cite{Liu:2012zya}. Further, $h_0$ can be fixed from the pion mass dependence of the $D$ meson masses.

In what follows, we will also consider $\bar{B} \to D$ transitions with the emission of two light
pseudoscalars (pions). Here,  chiral symmetry  puts constraints on one of the two pions while the
other one moves fast and does not participate in the final-state interactions. The corresponding
chiral effective Lagrangian has been developed in Ref.~\cite{Savage:1989ub}:
\begin{eqnarray}\label{LBDpipi}
{\cal L}_{\rm eff} &=& \bar B \big[ c_1 \left(u_\mu t M + Mtu_\mu \right)
+ c_2 \left(u_\mu M + Mu_\mu \right)t  + c_3\, t \left(u_\mu M+Mu_\mu \right)\nonumber\\
&+& c_4  \left(u_\mu \langle M t\rangle + M \langle u_\mu t\rangle \right)
+ c_5\, t \langle M u_\mu\rangle + c_6 \langle \left(Mu_\mu+u_\mu M\right)t\rangle \big] \partial^\mu D^\dag~,
\end{eqnarray}
in terms of the $B$-meson triplet $\bar{B}=(B^-, \bar{B}^0, \bar{B}_s^0)$, $M$ is the matter field for the
fast-moving pion and $t =uHu$ is a spurion field for Cabbibo-allowed decays,
\begin{equation}
H=\begin{pmatrix}
0 & 0 & 0 \\
1 & 0 & 0 \\
0 & 0 & 0
\end{pmatrix}~.
\end{equation}  
In the $B\to D\pi\pi$ decays that we will discuss later, only some combinations of the
LECs $c_i$ ($i=1,\ldots,6$) appear, see Sect.~\ref{sec:Dpipi}.

\subsection{Unitarization schemes}
\label{sec:uni}

\begin{figure}[h]
\centering
\includegraphics[width=0.98\textwidth]{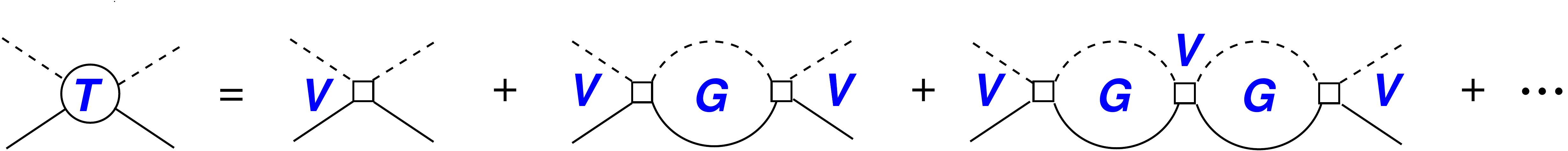}
\caption{The $T$-matrix for GB (dashed lines) scattering off $D$-mesons (solid lines) as a bubble sum based on the effective potential $V$, that is obtained from the underlying chiral effective Lagrangian.
\label{fig:uni}}
\end{figure}

As stated before, unitarity is only fulfilled perturbatively in CHPT. The hard scale in this EFT
is set by the appearance of resonances, like the $f_0(500)$ or the $\rho(770)$ in various partial
waves of pion-pion scattering. CHPT is thus not the proper framework to describe resonances. One
possible way to extend the energy region where this can be applied is unitarization, originally
proposed in Ref.~\cite{Dobado:1989qm}. However, this comes at a price, usually crossing symmetry
is violated in such type of approach and the coefficient of subleading chiral logarithms are
often incorrectly given, see~\cite{Gasser:1990bv}. Here, let us just discuss a familiar approach
on solving the Bethe-Salpeter equation in the on-shell approximation, see e.g~\cite{Oller:2000ma,Oller:2019opk}.
To be specific, let us consider the coupled-channel process $\phi+D \to \phi+D$ (suppressing all indices). The
basic unitarization method is depicted in Fig.~\ref{fig:uni}. It amounts to a resummation of the
so-called ``fundamental bubble'' (the 2-point loop function). To describe resonances like e.g.
the $D_{s0}^*(2317)$, one has to search for poles of the $T$-matrix, which is generated from the
CHPT potential by unitarization. 

\noindent This version of unitarized CHPT is based on the fundamental equation
\begin{equation}\label{eq:t}
T^{-1}(s) = V^{-1}(s) - G(s)~,
\end{equation}
where $V(s)$ is derived from the effective Lagrangian Eq.~(\ref{LDphi}) and $G(s)$ is the
2-point scalar loop function regularized by a subtraction constant ${a}(\mu)$,
\begin{eqnarray}\label{scalarloop}
G(s) &\!=&\! \frac{1}{16\pi^2}\bigg\{{a}(\mu)+\ln{\frac{m_2^2}{\lambda^2}}
+
\frac{m_1^2-m_2^2+s}{2s}\ln{\frac{m_1^2}{m_2^2}}
+\frac{\sigma}{2s}\left[\ln({s-m_1^2+m_2^2+\sigma}) \right. \nonumber\\
& & \left. -\ln({-s+m_1^2-m_2^2+\sigma}) +
\ln({s+m_1^2-m_2^2+\sigma})-\ln({-s-m_1^2+m_2^2+\sigma}) \right]
\bigg\},
\end{eqnarray}
with $\sigma=\left\{[s-(m_1+m_2)^2][s-(m_1-m_2)^2]\right\}^{1/2}$ and $m_1$ and $m_2$ are the
masses of the two mesons in the loop, here one $D$-meson and one GB. $\mu$ is
the scale of dimensional regularization, and a change of $\mu$ can be absorbed by
a corresponding change of ${a}(\mu)$. Promoting $T(s)$, $V(s)$ and $G(s)$ to be matrix-valued
quantities, it is easy to  generalize Eq.~(\ref{eq:t}) to coupled channels. More details on the
unitarization schemes will be given in the subsequent sections.

\subsection{Unitarized chiral perturbation theory in a finite volume}
\label{sec:FV}

To compare with lattice data, we need to formulate the unitarization scheme in a finite volume (FV).
Obviously, in any FV scheme, momenta are no longer continuous but quantized,
\begin{equation}
\vec q = \displaystyle\frac{2\pi}{L} \vec{n}~,~~\vec{n} \in \mathbb{Z}^3~,
\end{equation}
in a cubic volume of length $L$, i.e. $V = L^3$. An appropriate FV representation of the scalar 2-point
function is (see Ref.~\cite{Doring:2011vk} for details)
\begin{equation}
  \tilde{G}(s,L) =  \displaystyle\lim_{\Lambda\to\infty}\left[ \frac{1}{L^3} \sum_{\vec n}^{|\vec{q}|<\Lambda}
    I(\vec{q}\,) - \int_0^\Lambda \frac{q^2 dq}{2\pi^2} I(\vec{q}\,) \right]~,
\end{equation}
with $ I(\vec{q}\,)$ the corresponding integrand. The  FV energy levels of the process under consideration are then
obtained from the  poles of $\tilde{T} (s,L)$:
\begin{equation}
\label{eq:tFV}
\tilde{T}^{-1}(s,L) = V^{-1}(s) - \tilde{G}(s,L)~.
\end{equation} 
Note that all volume dependence resides in $\tilde{G}(s,L)$, the effective Lagrangian and thus
the effective potential are the same as in the continuum~\cite{Gasser:1987zq}. Again, in case
of coupled channels, Eq.~(\ref{eq:tFV}) is promoted to a matrix equation.

\section{The story of the $\Lambda(1405)$}
\label{sec:L1405}

\subsection{Basic facts}

In the quark model, the $\Lambda(1405)$ is a $uds$ excitation with $J^P = 1/2^-$ a few hundred
MeV above the ground-state $\Lambda(1116)$. The RPP gives {\bf one} corresponding state with
\begin{equation}
m=1405.1^{+1.3}_{-1.0}~{\rm MeV}\, , ~\Gamma = 50.5\pm 2.0\,{\rm MeV}~.
\end{equation}
\begin{wrapfigure}{r}{0.4\textwidth}
\centering
\vspace*{-.1cm}
\includegraphics[width=0.37\textwidth]{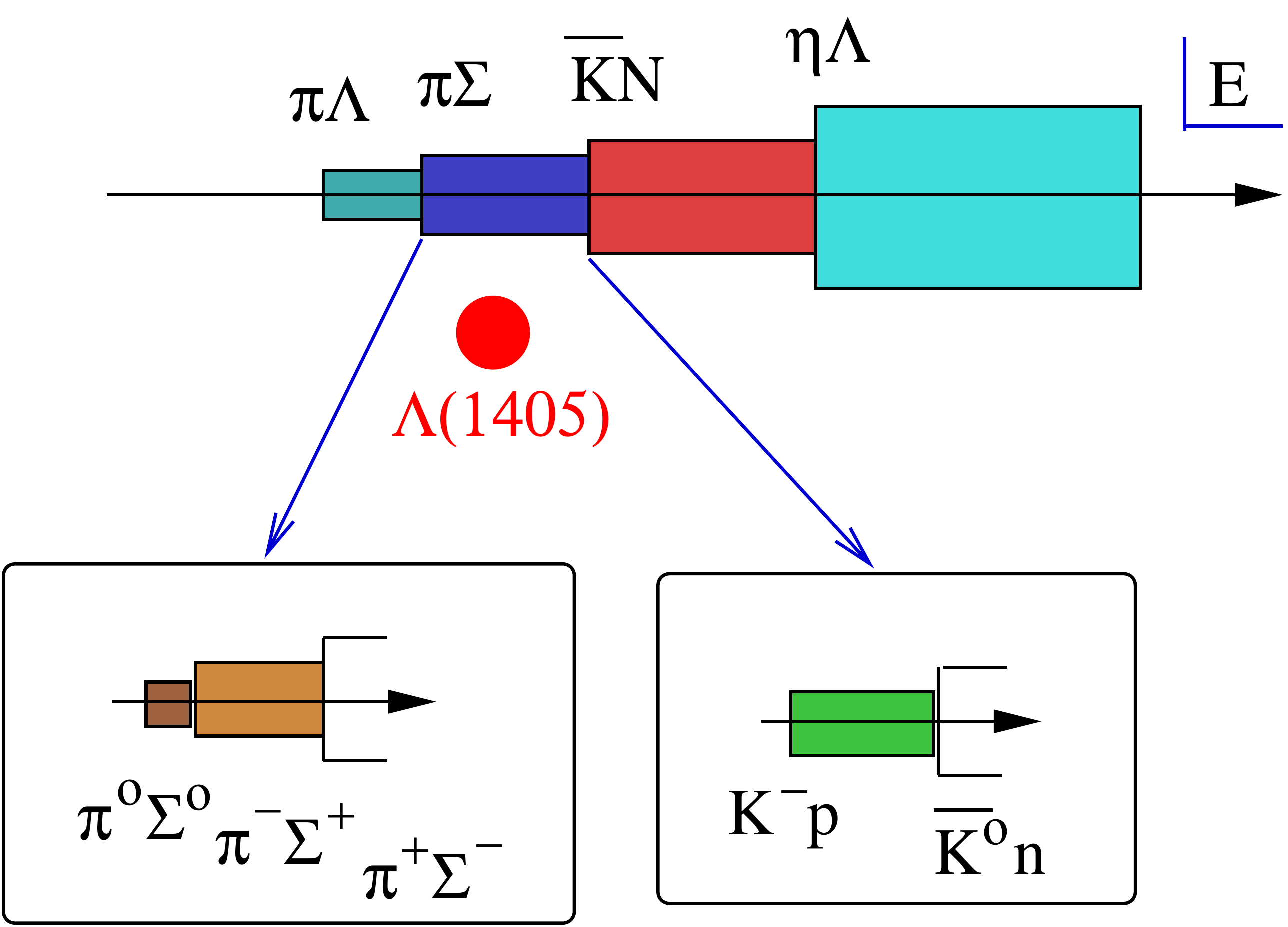}
\caption{Complex energy plane in the vicinity of the $\Lambda(1405)$.\label{fig:cuts}}
\end{wrapfigure}
In fact, the $\Lambda(1405)$ was predicted long before the quark model as a resonance in the
coupled $\pi\Sigma$  and $\bar{K}N$ channels, see~\cite{Dalitz:1959dn} and also~\cite{Kim:1965zzd},
and considered as a $\bar{K}N$ bound state, arguably the first ``exotic'' hadron ever. The analytical
structure in the complex energy plane between the $\pi\Lambda$ and the $\eta\Lambda$ thresholds
is shown in Fig.~\ref{fig:cuts}, together with the location of the the $\Lambda(1405)$ and the
further isospin splitting of the pertinent $\pi\Sigma$ and $\bar{K}N$ thresholds. The $\Lambda(1405)$ was clearly
seen in $K^-p\to \Sigma 3\pi$ reactions at 4.2~GeV at CERN~\cite{Hemingway:1984pz}. The spin and parity
were only recently determined directly in photoproduction reactions at Jefferson Laboratory, consistent
with the theoretical expectation of  $J^P = 1/2^-$~\cite{Moriya:2014kpv}. However, it is too
low in mass for the quark model, but can be described in certain models like the cloudy bag model\footnote{It is
  amusing to note that the two-pole structure of the $\Lambda(1405)$ was already observed in this
  model but little attention was paid to this work~\cite{Fink:1989uk}.}
or the Skyrme model. However, these models are only loosely rooted in QCD and do not allow for controlled
error estimates, an important ingredient in any theoretical prediction.

\smallskip

\subsection{Enter chiral dynamics}

An important step in the theory of the $\Lambda(1405)$  was based on the idea of combining the (leading order)
chiral SU(3) meson-baryon Lagrangian  with coupled-channel dynamics~\cite{Kaiser:1995eg}. This study
gave an excellent description of the  $K^-p \to K^- p, K^0n, \pi^0\Lambda, \pi^\pm\Sigma^\mp, \pi^0\Sigma^0$
scattering data and the $\pi\Sigma$ mass distribution, and it was found that the Weinberg-Tomozawa (WT) term
gave the most important contribution. In this scheme, the $\Lambda(1405)$ appears as a {\bf dynamically
  generated state}, a meson-baryon molecule, connecting the pioneering coupled-channel works
with the chiral dynamics of QCD. This led to a number of highly cited follow-up papers by the groups
from Munich and Valencia, see e.g. Refs.~\cite{Kaiser:1996js,Oset:1997it} and the early review~\cite{Oller:2000ma}.
These groundbreaking works were, however, beset by certain shortcomings. In particular, there was an
unpleasant regulator dependence for the employed Yukawa-type functions or momentum cutoffs and the issue
of maintaining gauge invariance in such type of regulated theories was only resolved years
later~\cite{Nacher:1999ni,Borasoy:2005zg,Borasoy:2007ku}.

\subsection{The two-pole structure}

A re-analysis of coupled-channel $K^- p$ scattering and the $\Lambda(1405)$ in the framework of
unitarized CHPT was performed in Ref.~\cite{Oller:2000fj}. This work was originally motivated by developing
methods to overcome some of the shortcomings discussed before. The following technical improvements
were worked out: 1) The subtracted meson-baryon loop function based on dimensional regularization, cf.
Eq.~(\ref{scalarloop}), which has become the standard regularization method; 2) A coupled-channel approach
to the $\pi\Sigma$ mass distribution, which replaced the common assumption that this process is dominated
by the $I=0$ $\pi\Sigma$ system and thus can be calculated directly from the $\pi\Sigma\to\pi\Sigma$ S-wave
amplitude; 3)  Matching formulas to any order in chiral perturbation theory were established, which allows
for a better constraining of such non-perturbative amplitudes.
The most significant finding of that work was,
however, the finding of the two-pole structure: ``Note that the $\Lambda(1405)$ resonance is described by
{\bf two poles} on sheets II and III with rather different imaginary parts indicating a clear departure from
the Breit-Wigner situation.''
\begin{wrapfigure}{r}{0.57\textwidth}
\centering
\vspace*{-.1cm}
\includegraphics[width=0.57\textwidth]{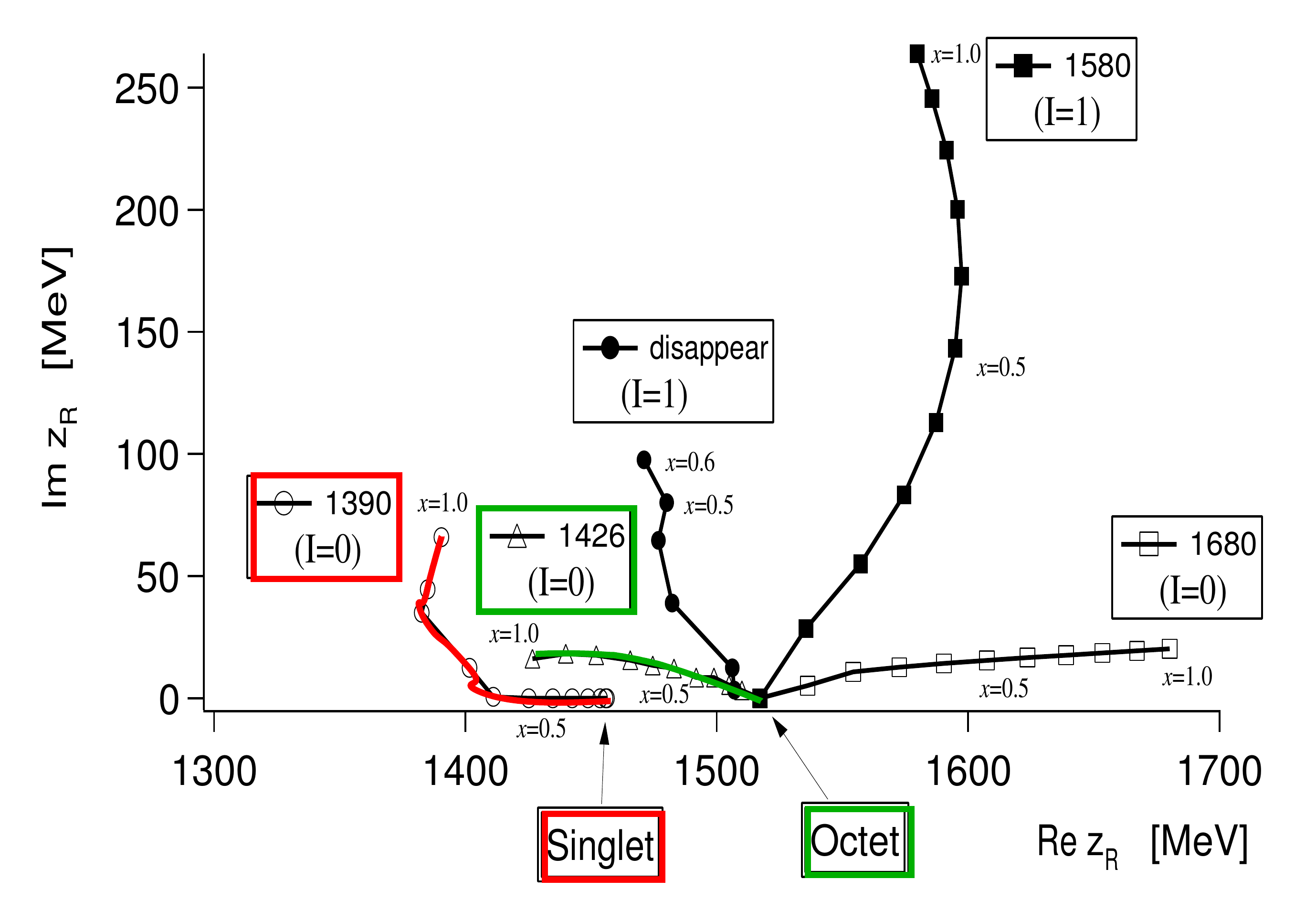}
\caption{Trajectories of the poles in the scattering amplitudes obtained by changing the SU(3) breaking
  parameter $x$. In the SU(3) limit ($x=0$), only two poles appear, one for the singlet and the
  other for the octets. The symbols correspond to the step size $\delta x=0.1$ and the two trajectories
  contributing to the $\Lambda(1405)$ are high-lighted.
  \label{fig:su3traj}}
\vspace{-5mm}
\end{wrapfigure}
The location of the poles are: Pole~1 at $(1379.2 -i 27.6)~{\rm MeV}$ and pole~2
at $(1433.7 -i 11.0)~{\rm MeV}$ on sheet~II, close to the $\pi\Sigma$ and $K^-p$ thresholds, respectively.  This two-pole
structure was also found in follow-up works by other groups~\cite{Jido:2002yz,GarciaRecio:2002td}.
A better understanding of this two-pole structure was achieved in Ref.~\cite{Jido:2003cb}
using  SU(3) symmetry considerations and group theory. For the case under consideration,
namely the dynamical generation of resonances in Goldstone boson scattering off baryons,
the following group theoretical consideration applies: The decomposition of the combination of the
two octets, the Goldstone bosons and the ground-state baryons, is
\begin{equation}
8 \otimes 8 = \underbrace{1 \oplus 8_s \oplus 8_a}_{\rm binding~at~LO}   \oplus 10  \oplus  \overline{10}  \oplus 27~,
\end{equation}
where using the leading order WT-term one finds poles in the singlet and in the two octets.
The two octets are degenerate and the poles are located on the real axis, see Fig.~\ref{fig:su3traj}.
One can now follow the developments of these poles in the complex plane. For that, one parameterizes
the departure from the SU(3) limit for the GB ($M_i$) and baryon masses ($m_i$) and subtraction constants
$a_i$ (the subtraction
  constants can be channel-dependent but collapse to one value in the SU(3) limit) as follows:
\begin{equation}
  M_i^2(x) = M_0^2 + x(M_i^2-M_0^2)~,~~
  m_i(x) = m_0 + x(m_i-m_0)~,~~
  a_i(x) = a_0 + x(a_i-a_0)~,
\end{equation}
with $0\leq x\leq 1$, where $x = 0$ corresponds to the SU(3) limit and $x =1$ describes the physical world. 
Further, $m_0=1151\,$MeV, $M_0=368\,$MeV and $a_0=-2.148$. The trajectories of the various poles in the complex
plane as the SU(3) breaking is gradually increased up to the physical values at $x=1$ is shown in
Fig.~\ref{fig:su3traj}. First, we observe that not all poles present in the SU(3) limit appear for $x=1$ (using the
LO WT term only).
Second, what concerns the $\Lambda(1405)$, we see that the singlet pole moves towards the $\pi\Sigma$
threshold and becomes rather broad, whereas the second pole from the octet comes out close to the $K^-p$ threshold
and stays rather narrow. So there are in fact two resonances. Having determined these poles, one can determine
the couplings of these resonances to the physical states by studying the amplitudes in the vicinity of the
poles,
\begin{equation}
T_{ij} = \frac{g_ig_j}{z-z_R} + {\rm regular~terms}~,
\end{equation}    
with $i,j$ channel indices and the couplings $g_i$ are complex valued numbers. While the lower pole
couples stronger to the $\pi\Sigma$ channel, the higher one displays a stronger coupling to $\bar{K}N$.
Consequently,  it is possible to find  the existence of the two resonances by performing different
experiments, since in different experiments the weights by which the two resonances are excited are different,
see Ref.~\cite{Jido:2003cb} for more details. Further early support of the two-pole scenario was provided by the leading
order investigation of the reaction $K^-p\to \pi^0\pi^0\Sigma^0$~\cite{Magas:2005vu}.

\subsection{Beyond leading order}

\begin{wrapfigure}{r}{0.45\textwidth}
\centering
\vspace*{-.1cm}
\includegraphics[width=0.44\textwidth]{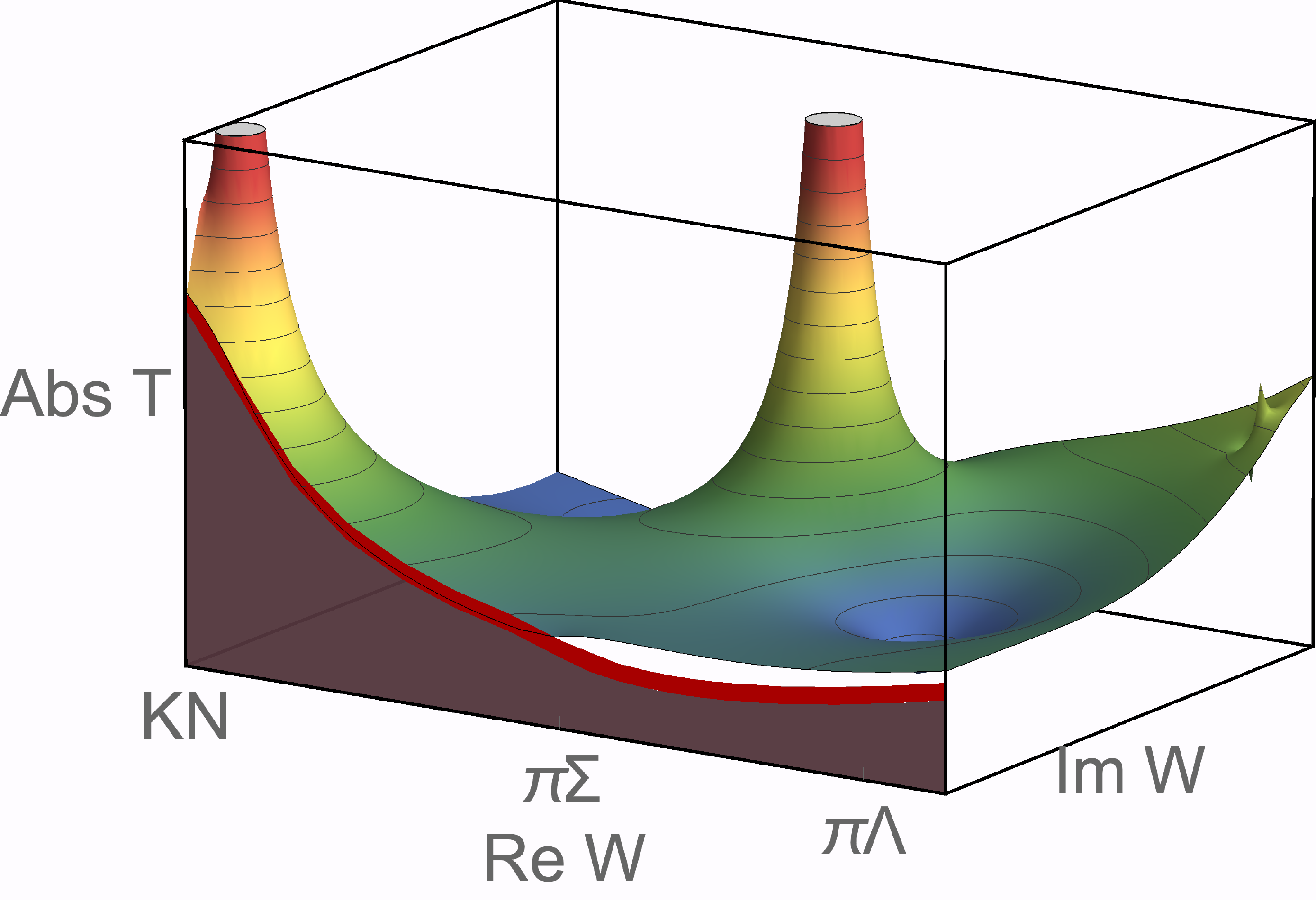}
\caption{3-dimensional plot of the two-pole structure of the $\Lambda(1405)$. $W$ denotes the center-of-mass
  energy. Figure courtesy of Maxim Mai.
  \label{fig:twopole}}
\vspace{-5mm}
\end{wrapfigure}
Clearly, to achieve a better precision, one has to go beyond leading order and include the next-to-leading
order (NLO) terms in the chiral SU(3) Lagrangian in the effective potential. This task was performed
by three groups independently~\cite{Ikeda:2012au,Guo:2012vv,Mai:2012dt}. These investigations were also
triggered by the precise measurements of the energy shift and width of kaonic hydrogen~\cite{Bazzi:2011zj},
which was based on the improved Deser-type formula from Ref.~\cite{Meissner:2004jr}, thus resolving the
long standing ``kaonic hydrogen puzzle'' (the discrepancy between the values of the $\bar{K}N$ scattering
lengths extracted from scattering data or earlier kaonic hydrogen experiments). This allowed to pin
down the subthreshold $K^-p$ scattering amplitude to better precision and to make predictions for the 
$K^-n$ scattering lengths. Most importantly, all of these works confirmed the two-pole structure as
shown in Fig.~\ref{fig:twopole} from~\cite{Mai:2012dt}.

\begin{figure}[t]
\centering
\includegraphics[width=0.98\textwidth]{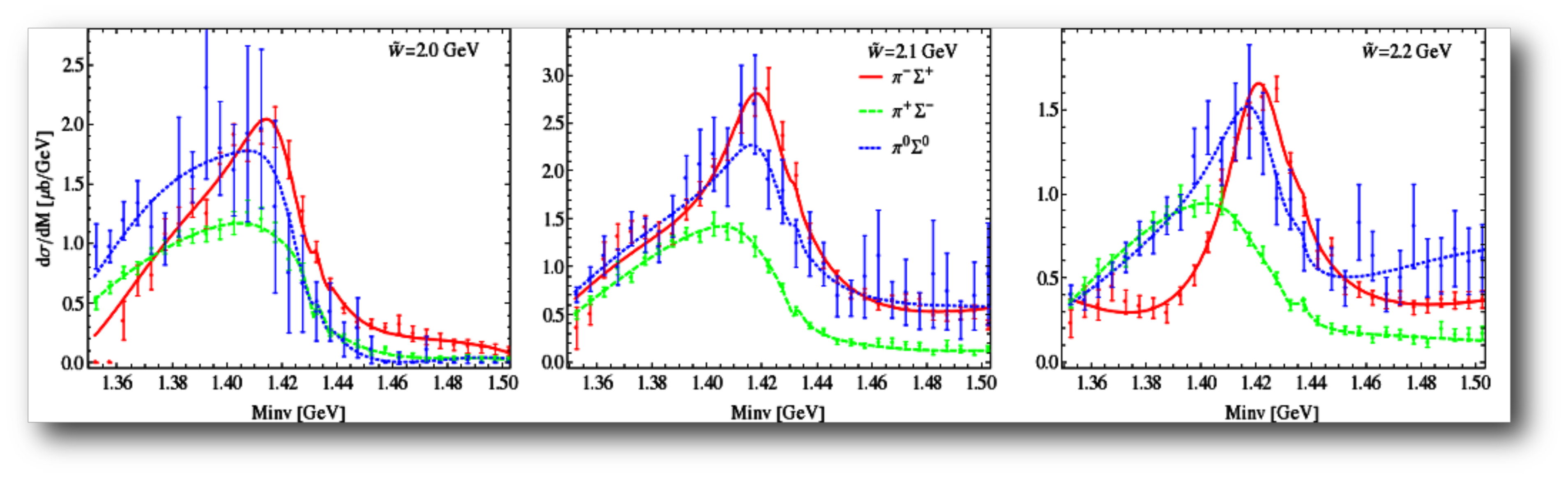}
\caption{Result of the fits to the CLAS data in all three channels $\pi^+\Sigma^-$ (green), $\pi^-\Sigma^+$(red)
  and $\pi^0\Sigma^0$. Correspondingly, green (dashed), red (full) and blue (dotted) lines represent the outcome
  of the model for the solution~4 in the $\pi^+\Sigma^-,\pi^-\Sigma^+,\pi^0\Sigma^0$  channels, respectively.
  Figure courtesy of Maxim Mai.
  \label{fig:photo}}
\end{figure}

There was yet another surprise, namely by looking more closely at the scattering and kaonic hydrogen data, one
can find at least eight solutions of similar quality with different pairs of poles for the $\Lambda(1405)$,
see Ref.~\cite{Mai:2014xna}. Here, photoproduction comes to the rescue. The CLAS collaboration at
Jefferson Laboratory did a superb job in measuring  the $\Sigma\pi$ photoproduction line shapes near the
$\Lambda(1405)$~\cite{Moriya:2013eb}. These data were first analyzed using LO unitarized CHPT and a 
polynomial ansatz for the photoproduction process $\gamma p \to K^+M_iB_i$ in Ref.~\cite{Roca:2013av}, leading to a good
fit of these data and a further check of the two-pole nature of the $\Lambda(1405)$. The same ansatz
was used in~\cite{Mai:2014xna}, leaving only two of the eight solutions, as shown for one of the remaining
solutions in Fig.~\ref{fig:photo} for a fraction of the data (the complete fit is shown in~\cite{Mai:2014xna}). 
Similarly, the spread in the two poles from the eight solutions was sizeably reduced, although the lower
pole could not be pinned down as well as the higher one. This work also supplied error bands, not only
for the photoproduction results but also for the underlying hadronic scattering processes. This shows that
the inclusion of the photoproduction data serves as  a new important constraint on the antikaon-nucleon
scattering  amplitude.   In view of all these NLO results, the two-pole structure first appeared
in the RPP in form of a mini-review~\cite{Patrignani:2016xqp}, called ``Pole structure of the $\Lambda(1405)$
region'' co-authored by Hyodo and Mei{\ss}ner, which give also references to other works related to the
two-pole scenario. Still, in the RPP summary tables the  $\Lambda(1405)$ was listed (and still is) as one
resonance only.

%

\subsection{Where do we stand?}

Low-energy meson-baryon interactions in the strangeness $S=-1$ sector have been considered by a number of groups
world-wide. A comparative study of various chiral unitary approaches was given in  Ref.~\cite{Cieply:2016jby}.
These approaches are based on a chiral Lagrangian that includes terms up to NLO, ${\cal O}(p^{2})$
\cite{Ikeda:2012au,Guo:2012vv,Mai:2012dt,Cieply:2011nq}. The various LECs (and others parameters) are
determined from a fit to the low-energy $K^{-}p$ scattering data as well as to the properties of  kaonic hydrogen 
measured  by the SIDDHARTA collaboration. It was thus possible to not only compare the pole content of these
various approaches but also the subthreshold energy dependence of the $\bar{K}N$ scattering amplitudes,
which are important e.g. in the description of kaonic nuclei. These models differ in technical details
but also in some underlying philosophical aspects. Largely, dimensional regularization
is used to tame the ultraviolet  divergences in the meson-baryon  loop function
and the meson-baryon interactions are considered on the energy shell. This holds for the  Kyoto-Munich, Murcia and
Bonn models. Differently from that, the Prague model introduces off-shell form factors to regularize the Green 
function and phenomenologically accounts for the off-shell effects. In all but one of these approaches
an  effective meson-baryon potential is introduced and matched to  the chiral amplitude up to a given order.
This potential is then iterated in a Lippmann-Schwinger equation to generate the bound and scattering states.
\begin{wrapfigure}{r}{0.45\textwidth}
\centering
\vspace*{-.1cm}
\includegraphics[width=0.44\textwidth]{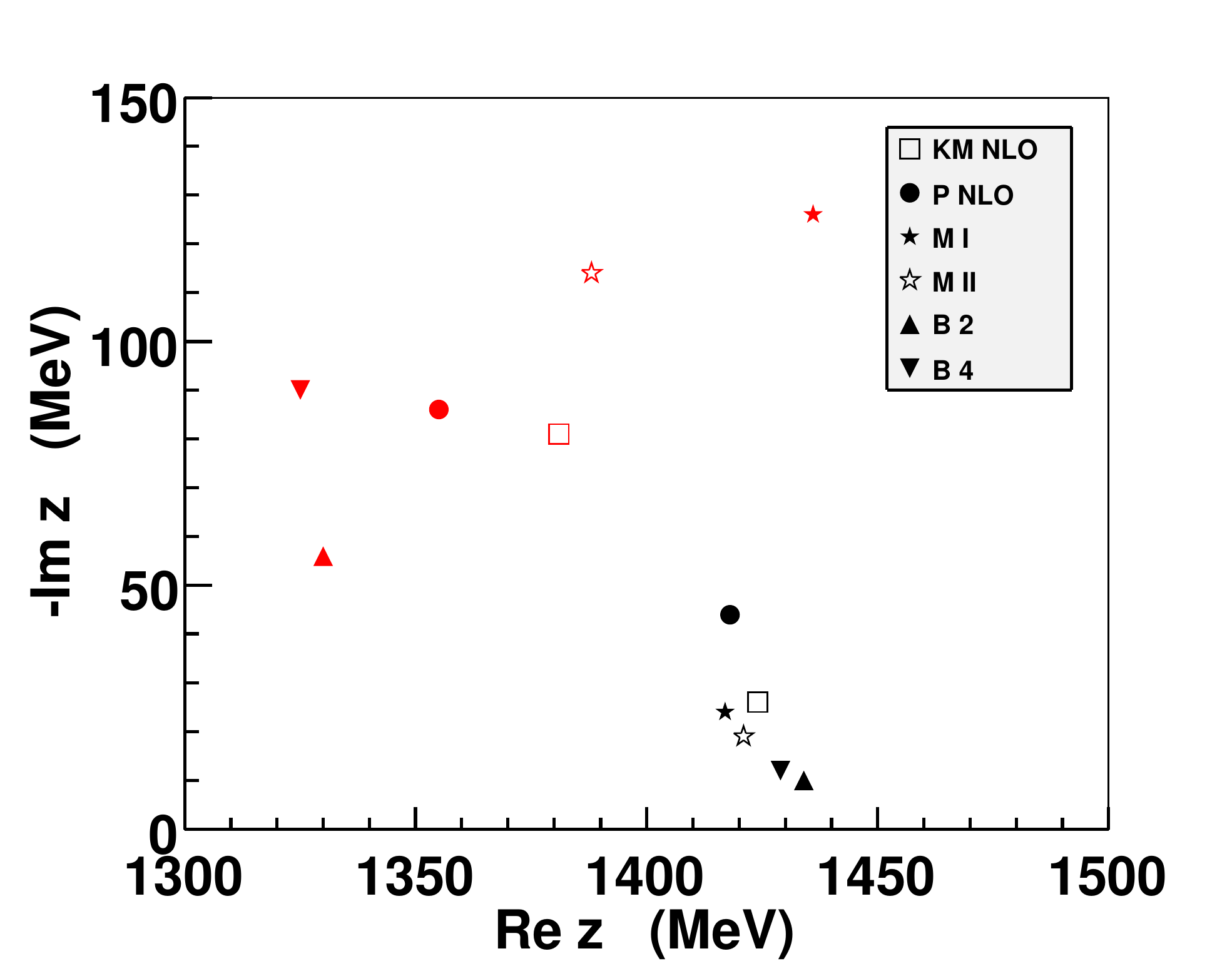}
\caption{Pole positions for various approaches: Kyoto-Munich (KM)~\cite{Ikeda:2012au},
  Prag (P)~\cite{Cieply:2011nq}, Murcia (M)~\cite{Guo:2012vv} and Bonn (B)~\cite{Mai:2012dt}.
  \label{fig:poles}}
\vspace{-3mm}
\end{wrapfigure}
In case of the  Bonn model, a  Bethe-Salpeter  equation is solved followed by an S-wave projection. Also, the
off-shell contributions are neglected. Further, in the  Kyoto-Munich and Prague approaches, the NLO corrections
to the LO part are kept moderately small while in the Murcia and Bonn models sizable NLO terms  are generated.
These lead to inter-channel couplings very different from those obtained by only  the WT interaction.
Nevertheless,  these different  models reproduce the experimental data on 
a qualitatively very similar level. They are also in mutual agreement for the data 
available at the $\bar{K}N$ threshold and they agree on the position of the 
higher pole for the $\Lambda(1405)$ resonance. The  predicted 
complex energy has a real part {$\Re z \approx 1420\ldots1430$~MeV} and an imaginary part 
{$-\Im z \approx 10\ldots40$~MeV}, see Fig.~\ref{fig:poles}.
These different approaches, however, also lead to some distinct differences in their predictions.
In particular, one finds very different locations of the lighter pole and different predictions for the
elastic $K^{-}p$ and $K^{-}n$ amplitudes
at sub-threshold energies. This certainly has impact on the predictions one can make (or has made)
for kaonic atoms and antikaon quasi-bound states, see e.g.~\cite{Mares:2014tfa}.

Clearly, we need better data to constrain the $\pi\Sigma$ spectra in various processes besides the
already mentioned photoproduction data. A step in this direction was performed in Ref.~\cite{Kamiya:2016jqc}.
These authors provide further constraints on the $\bar{K}N$ subthreshold interaction  by analyzing
$\pi\Sigma$ spectra in various processes, such as the $K^-d \to \pi \Sigma n$ reaction and the $\Lambda_c \to
\pi \pi \Sigma$ decay. Also, the yet to be measured 1S level shift of kaonic deuterium will put further
constraints on the $\bar{K}N$ interaction~\cite{Meissner:2006gx,Hoshino:2017mty}. For the status of
such measurements, see~\cite{Curceanu:2020kkg}. Further progress has also been made by including the
P-waves and performing a sophisticated error analysis~\cite{Sadasivan:2018jig}, confirming again
and sharpening further the two-pole scenario of the $\Lambda(1405)$ (note that the
P-waves had already been included earlier with a focus on the $S=0$ sector~\cite{CaroRamon:1999jf}).
Despite these remaining uncertainties, it  must be stated clearly that the two-pole nature of
the $\Lambda(1405)$ is established!

To end this section, I briefly discuss two recent works that challenged the two-pole structure. It was
claimed in Ref.~\cite{Revai:2017isg} that the two-pole nature is an artifact of the on-shell approximation
used in most studies. Including off-shell effects, only one pole is generated in that study. However,
as convincingly shown in Ref.~\cite{Bruns:2019bwg}, that approach violates constraints imposed by
chiral symmetry. What is the origin of this violation? It is partly due to the treatment of the off-shell 
chiral-effective vertices, in combination with the employed regularization scheme and the use of the non-relativistic
approximation. Overcoming these deficiencies, the two-pole scenario reappears. This does not come
as a surprise as the NLO study of Ref.~\cite{Mai:2012dt} already went beyond the on-shell approximation.
Another recent paper with only one pole is Ref.~\cite{Anisovich:2019exw}, based on the phenomenological
Bonn-Gatchina  (BnGa) approach. In this framework, a large number of scattering and photoproduction data is
fitted. However, this scheme does not allow for the dynamical generation of resonances and no pole searches in the
complex-energy plane are reported in~\cite{Anisovich:2019exw}.  In a recent update, these authors also included a second pole in their model~\cite{Anisovich:twopole}.  They find that the second pole does not considerably worsen the description of the considered data but still they prefer the solution with one pole. Thus, these results are not conclusive.
In addition, data that further support the two-pole nature on $\pi^-p\to K^0\pi\Sigma$ and $pp\to pK^+\pi\Sigma$
\cite{Bayar:2017svj} as well as $K^+\Lambda(1405)$ electroproduction~\cite{Lu:2013nza} 
are also not included. It can therefore safely be said that these papers do indeed {\em not}
challenge the two-pole scenario.

\section{Meson sector:  The $D_0^*(2300)$ and related states}
\label{sec:D2300}

So far, one might consider the two-pole structure a curiosity related to just one particular state.
In the meson sector, a similar doubling of states was already considered in 1986 in the discussion
of the mysterious decay properties of the $f_0(980)$ (then called $\delta(980)$), but that remained
largely unnoticed~\cite{Cahn:1985wu}. However,
let us now take a closer look at the spectrum  of the excited charmed mesons, especially the
$D_{s0}^*(2317)$ first observed by  BaBar~\cite{Aubert:2003fg} and the $D_0^*(2300)$ first observed
by Belle~\cite{Abe:2003zm} (see also Ref.~\cite{Link:2003bd}). According to the recent edition of the
PDG, the characteristics of these states are:
\begin{eqnarray}
D_0^*(2300)&:&~~ M = 2300\pm 19~{\rm MeV}~,~~\Gamma = 274\pm 40~{\rm MeV}~,~~I(J^P)=\frac{1}{2}(0^+)~,
\label{D0}\\
D_{s0}^*(2317)&:&~~ M = 2318.0\pm 0.7~{\rm MeV}~,~~\Gamma < 3.8~{\rm MeV}~,~~I(J^P)=0(0^+)~.
\label{Ds0}
\end{eqnarray}  
According to the quark model, the quark composition for these scalar mesons is $c\bar{u}$ and $c\bar{s}$,
respectively. This immediately poses the question: Why is the $D_{s0}^*(2317)$ as heavy as the $D_0^*(2300)$,
it should be about 100~MeV, which is the mass of the strange quark, heavier? Also, why is  the $D_{s0}^*(2317)$
about 150~MeV
below the prediction of the quark model, that has been rather successful~\cite{Godfrey:1985xj}? 
While this is an interesting question
(I refer to the review~\cite{Guo:2017jvc} which has a very detailed discussion of this state), 
here I will focus on the the non-strange charmed scalar meson, as it appears to be too heavy, but in fact
will give further support to the two-pole scenario. 

\subsection{Two-pole structure}

Let us consider first the fine lattice QCD work by the Hadron Spectrum Collaboration, who investigated
coupled-channel $D\pi$, $D\eta$ and $D_{s}\bar{K}$ scattering with $J^P=0^+$ and $I=1/2$ in three lattice volumes,
one value for the temporal and the spatial lattice spacing, respectively,  at a pion mass
$M_\pi=391\,$MeV and $D$-meson mass $M_D = 1885\,$MeV \cite{Moir:2016srx}. They used various $K$-matrix
type extrapolations of the type
\begin{equation}\label{KM}
  K_{ij} = \left(g_i^{(0)}+g_i^{(1)}s\right) \left(g_j^{(0)}+g_j^{(1)}s\right)\frac{1}{m^2-s}
  +\gamma_{ij}^{(0)} + \gamma_{ij}^{(1)}s~,
\end{equation}    
to find the poles in the complex plane, by fitting the parameters $g, \gamma$ to the computed energy
levels, and use the $T$-matrix to extract the poles. They found one S-wave pole at $(2275.0\pm0.9)\,$MeV,
extremely close to the $D\pi$ threshold at 2276~MeV. This state is consistent with the $D_0^*(2300)$ of the PDG.
However, the extrapolations in Eq.~(\ref{KM}) do not take into account chiral symmetry.
\begin{figure}[t]
\centering
\vspace*{-.1cm}
\includegraphics[width=0.45\textwidth]{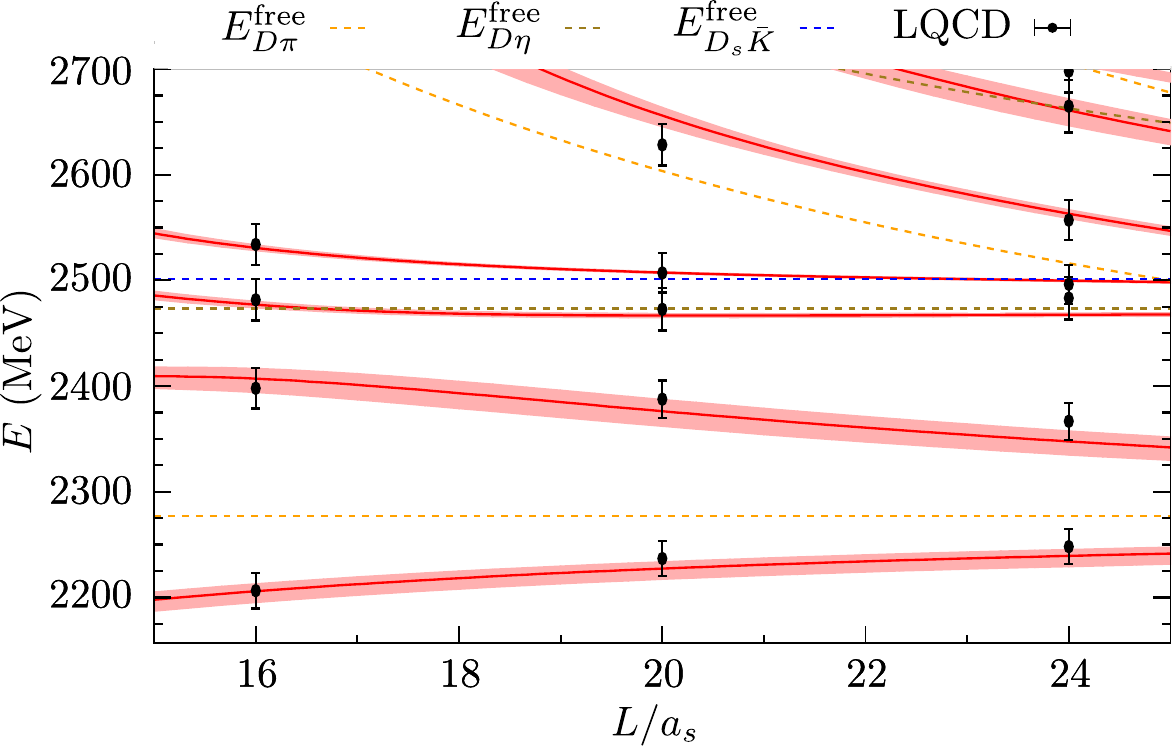}~~~
\includegraphics[width=0.45\textwidth]{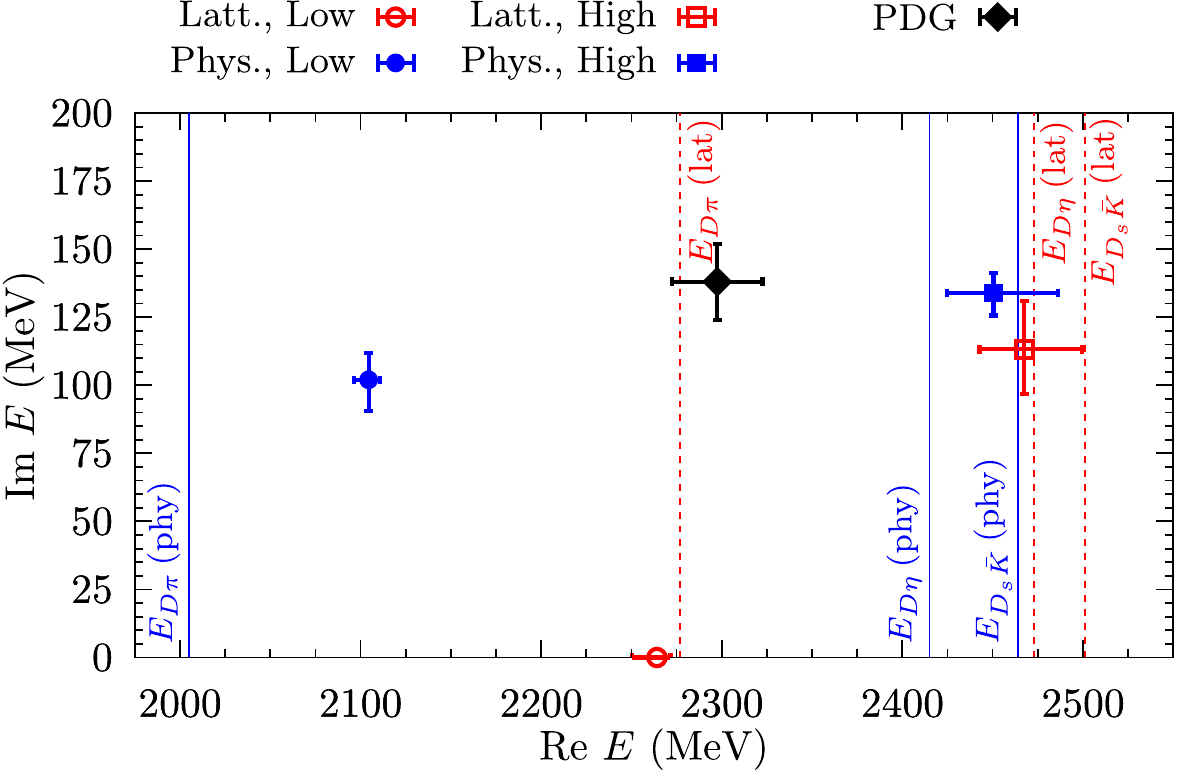}~~~
\caption{Left panel: Energy levels calculated in finite-volume unitarized chiral perturbation theory
  with all LECs determined before (red bands representing the 1$\sigma$ uncertainties) in comparison to
  the lattice QCD results of~\cite{Moir:2016srx} (black circles). The dashed lines give the various
  free levels of the two-particle systems $D\pi$, $D\eta$ and $D_s\bar{K}$. Right panel: Location of the two poles
  in the complex energy plane for the lattice masses (red symbols) and physical masses (blue symbols).
  The black diamond represents the PDG value. The various thresholds are indicated by the dotted lines.
  Figures courtesy of Feng-Kun Guo.}
\label{fig:D0levelspoles}
\vspace{-.3mm}
\end{figure}

Therefore, this topic was revisited in Ref.~\cite{Albaladejo:2016lbb}, where the chiral Lagrangian, Eq.~(\ref{LDphi}),
together with LECs from Ref.~\cite{Liu:2012zya} was implemented within the finite volume formalism
outlined in Sect.~\ref{sec:FV} to postdict in a parameter-free manner the energy levels measured by
the Hadron Spectrum Collaboration. The stunning result of this unitarized CHPT calculation is shown in
the left panel of Fig.~\ref{fig:D0levelspoles}, a very accurate postdiction of the lattice levels is
achieved (note that this is not a fit). Note further that the region above 2.7~GeV is beyond the range of applicability
of this NLO calculation. The level below the $D\pi$ threshold is interpreted in Ref.~\cite{Moir:2016srx} as
a bound state associated to the  $D_0^*(2300)$ as stated before. The finite-volume UCHPT calculation also
finds this pole at $M = 2264_{-14}^{+8}\,$MeV and half-width $\Gamma/2=0\,$MeV, very similar to the results
of  Ref.~\cite{Moir:2016srx}. However, there is a {\bf second pole} at $M= 2468_{-25}^{+32}\,$MeV with 
$\Gamma/2=113^{+18}_{-16}\,$MeV, see also the right panel of Fig.~\ref{fig:D0levelspoles}. Using chiral
extrapolations, one can then evaluate the spectroscopic content of the scattering amplitudes for  the physical pion mass,
collected in Tab.~\ref{tab:D0}.

\renewcommand{\arraystretch}{1.3}

\begin{table}[htb!]
  \caption{Position $\sqrt{s}=M-i\Gamma/2$ (in MeV) and couplings (in GeV) of the two poles in
  the (0,1/2) sector using physical pion masses.\label{tab:D0}}
\centering
\vspace{+2mm}
\begin{tabular}{|ccccc|}
\hline
$M$~(MeV) & $\Gamma/2$ (MeV) & $|g_{D\pi}|$ &  $|g_{D\eta}|$ &  $|g_{D_s\bar{K}}|$\\
\hline
$2105^{+6}_{-8}$  & $102^{+10}_{-12}$  & $9.4^{+0.2}_{-0.2}$ & $1.8^{+0.7}_{-0.7}$ & $4.4^{+0.5}_{-0.5}$\\
$2451^{+36}_{-26}$  & $134^{+7}_{-8}$  & $5.0^{+0.7}_{-0.4}$ & $6.3^{+0.8}_{-0.5}$ & $12.8^{+0.8}_{-0.6}$\\
\hline
\end{tabular}
\end{table}
The bound state below the $D\pi$ threshold evolves into a resonance above it when the physical masses are
used, where the threshold is now at 2005~MeV. This behaviour is typical for S-wave poles. The second pole moves
very little and its couplings are rather independent of the meson masses. It is a resonance located between
the $D\eta$ and $D_s\bar{K}$ thresholds on the (110) Riemann sheet, continuously connected to the phyical sheet.
Thus, the $D_0^*(2300)$ of the RPP is produced by two different poles, and in fact the lower pole solves the
enigma discussed in the beginning of this section. Note that this two-pole structure was observed earlier
in Refs.~\cite{Kolomeitsev:2003ac,Guo:2006fu,Guo:2009ct} but only explained properly in ~\cite{Albaladejo:2016lbb}
as discussed next.

\begin{figure}[t!]
\centering
\vspace*{-.1cm}
\includegraphics[width=0.50\textwidth]{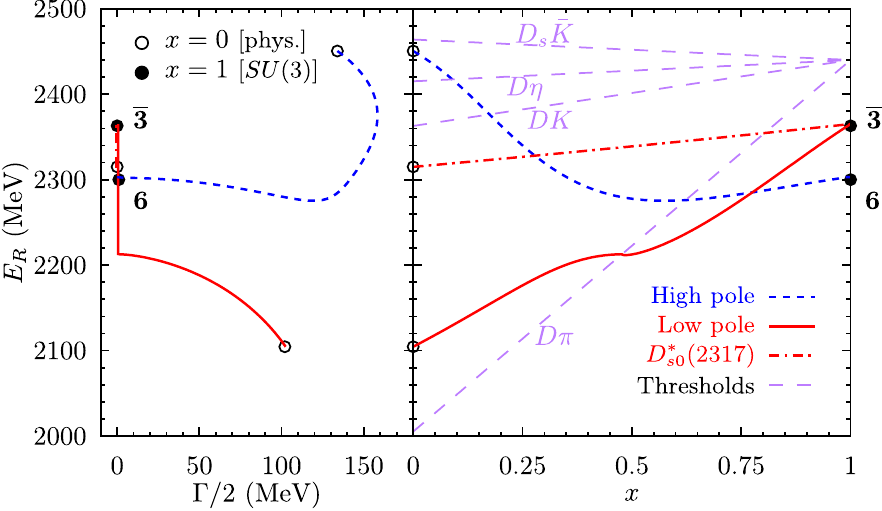}~~
\includegraphics[width=0.47\textwidth]{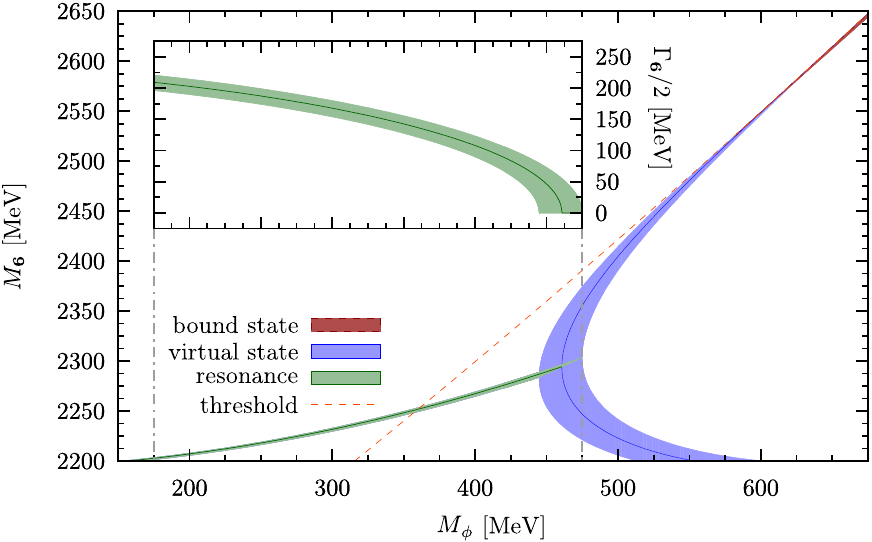}~~
\caption{Left panel: Pole paths in the complex plane when recovering the SU(3) limit (left subpanel).
  Mass evolution of the different poles with $x$. Besides the two (0,1/2) poles, denoted as
  high and low (blue dashed and green solid lines), the evolution of the (1,0) bound state, the
  $D_{s0}^*(2317)$ resonance (orange dot-dashed line), is shown (right subpanel). Figure courtesy of 
  Feng-Kun Guo. Right panel: Mass of the predicted sextet state $M_6$ at the SU(3) symmetric point as a function of the Goldstone boson mass $M_\phi$.
  The inset shows the half-width of the resonance for GB masses below 475~MeV.
  \label{fig:D0poles}} 
\vspace{-.3mm}
\end{figure}

Consider again the SU(3) limit, where the eight Goldstone bosons take the common value $M_{0}$ and the three
heavy $D$-mesons the common value $M_{D0}$, so that departures from the SU(3) limit are parameterized as
\begin{eqnarray}
  M_{\phi,i}  &=& M_{\phi,i}^{\rm phys} + x(M_0- M_{\phi,i}^{\rm phys})~(i=1,\ldots 8)~,\nonumber\\
  M_{D,j}  &=& M_{D,j}^{\rm phys} + x(M_{D0}- M_{D,j}^{\rm phys})~(j=1,\ldots 3)~,\nonumber
\end{eqnarray}  
with $x=0$ and $x=1$ corresponding to the physical and the SU(3) symmetric case, respectively, and 
$M_0=0.49\,$ GeV and $M_{D0}=1.95\,$GeV. In that study, only one subtraction constant for all channels
was used and kept fixed with varying $x$. Note that
in contrast to the work of Ref.~\cite{Jido:2003cb} discussed before, here a linear extrapolation formula
is used for the GB masses, which is also legitimate. As before,
the two-pole nature is understood from group theory, 
\begin{equation}
\bar{3}\otimes 8 = \underbrace{\bar{3} \oplus 6}_{\rm attractive} \oplus \overline{15}~.
\end{equation} 
This means one has attraction in the {\boldmath$\bar{3}$} and {\boldmath$6$} irreducible representations (irreps)
but repulsion in the {\boldmath$\overline{15}$} irrep at leading order in the effective potential.
The most attractive irrep, the {\boldmath$\bar{3}$}, admits a $c\bar{q}$
($q=u,d,s$) configuration. At NLO, the potentials receive corrections, but the qualitative features remain.
The evolution from the SU(3) limit to the physical case is shown in the left panel of Fig.~\ref{fig:D0poles}.
Shown are the two poles corresponding to the $D_0^*(2300)$ and its strange sibling, the $D_{s0}^*(2317)$.
As one moves away
from the SU(3) limit, the lower pole of the  $D_0^*(2300)$ moves down in the complex plane, restoring the
expected ordering that the $c\bar{u}$ excitation should be lighter than its $c\bar{s}$ partner.
It was already observed in Ref.~\cite{Albaladejo:2016lbb} that the higher pole connects with a virtual
state in the sextet representation due to the weaker binding.  This issue was further elaborated on in
Ref.~\cite{Du:2017zvv}, where the SU(3) limit was studied in more detail. In the right panel of Fig.~\ref{fig:D0poles}
the sextet pole is shown for varying GB masses. Below $M_\phi\lesssim475$~MeV, the pole is a resonance
with its imaginary part ($\Gamma_{\bm{6}}/2$) shown in the inserted sub-figure. Above $M_\phi\simeq 475$~MeV,
it  evolves into a pair of virtual states, and finally it becomes a bound state at $M_\phi\simeq 600$~MeV.
Such a study in the SU(3) limit $m_u=m_d=m_s$ at large GB masses could be performed on the lattice and
offer another test of this scenario. Finally, let me come back to the lattice calculation of
Ref.~\cite{Moir:2016srx}. Indeed, while various of the amplitudes employed in  that
analysis contained a second pole, its location was strongly parameterization-dependent~\cite{Davidprivate},
and therefore not reported in that paper.

\subsection{Other candidates}
As already noted, in the $(S,I)=(1,0)$ sector, the same chiral Lagrangian produces a pole at
$2315^{+18}_{-28}$~MeV which is naturally identified with the  $D_{s0}^*(2317)$. It emerges from the
pole in the triplet representation. The $D_{s0}^*(2317)$ is dominantly a $DK$ molecule. Substituting
the $D$-meson by a $D^*$ and employing HQSS, the molecular picture naturally gives~\cite{Cleven:2010aw}
\begin{equation}
M_{D_{s1}(2460)} - M_{D_{s0}^*(2317)} \simeq M_{D^*} - M_D~,
\end{equation}  
which is also obtained in the so-called doublet model~\cite{Bardeen:2003kt,Nowak:2003ra,Mehen:2005hc}.
Similar to the $D_0^*(2300)$, the nonstrange charmed meson $D_1$ also comes with two poles, see
Tab.~\ref{tab:poles}. This was noted before in \cite{Guo:2006rp}. In both cases the single RPP pole sits in between the lower and the higher
pole.

Using HQFS, one can further predict similar states in the $B$-meson sector,
by just replacing the corresponding $D$-mesons with their bottom counterparts at leading
order~\cite{Guo:2006fu,Cleven:2010aw}. Using the NLO framework employed in the charm sector, one
has to scale the LECs $h_{0,1,2,3} \sim m_Q$  and $h_{4,5}\sim 1/m_Q$ and the subtraction constant 
is adjusted as described in Ref.~~\cite{Guo:2006fu}. This leads again to the two-pole structures also
collected in Tab.~\ref{tab:poles}.

For the lowest positive-parity heavy strange mesons, it is instructive to compare with lattice QCD
results. This gives for the masses of the charm-strange mesons
$M_{D_{s0}^*} = 2315^{+18}_{-28}  [2348^{+7}_{-4}]$~\cite{Bali:2017pdv}~,
$M_{D_{s1}} = 2456^{+15}_{-21}  [2459.5\pm0.6]$~\cite{Bali:2017pdv}~,
and for the strange-bottom ones
$M_{B_{s0}^*} = 5720^{+16}_{-23}  [5711\pm23]$~\cite{Lang:2015hza}~,
$M_{D_{s1}} = 5772^{+15}_{-21}  [5750\pm25]$~\cite{Lang:2015hza}~, where the first [second] number refers
to the molecular [lattice QCD] prediction. The agreement is rather remarkable.

\renewcommand{\arraystretch}{1.3}

\begin{table}[htb!]
\caption{Predicted poles corresponding to the positive-parity heavy-light nonstrange mesons given
 as ($M,\Gamma/2$), with $M$ the mass and $\Gamma$ the  total decay width, in units of MeV. The current
 RPP~\cite{Tanabashi:2018oca} values are listed in the last column.}
\label{tab:poles}
\vspace{+2mm}
\centering
\begin{tabular}{|cccc|}
\hline
& \text{lower pole} & \text{higher pole} & \text{RPP} \\ 
\hline
$D_0^*$ & $\left(2105^{+6}_{-8}, 102^{+10}_{-11}\right)$
& $\left(2451^{+35}_{-26},134^{+7}_{-8}\right)$ & $(2300\pm19,137\pm20)$ \\
$D_1$ &  $\left(2247^{+5}_{-6}, 107^{+11}_{-10} \right)$  & $\left(2555^{+47}_{-30},
203^{+8}_{-9}\right)$ & $(2427\pm26\pm25,192^{+54}_{-38}\pm37)$
\\
$B_0^*$   & $\left(5535^{+9}_{-11},113^{+15}_{-17} \right)$ 
&  $\left(5852^{+16}_{-19},36\pm5\right)$ & - \\
$B_1$ &   $\left( 5584^{+9}_{-11}, 119^{+14}_{-17} \right)$ 
      &  $\left(5912^{+15}_{-18}, 42^{+5}_{-4}\right)$ & -\\ 
\hline
\end{tabular}
\end{table}

So the plot of the two-pole scenario thickens. In the absence of direct measurements of some of these
states, one might ask the question whether there is further experimental support for the picture
just outlined?

\subsection{Analysis of $B\to D\pi\pi$ data}
\label{sec:Dpipi}

To further test the mechanism of the dynamical generation of the charm-light flavored mesons
discussed so far, let me turn to the high precision results of the LHCb collaboration for the
decays $B\to D\pi\pi$~\cite{Aaij:2016fma}. As shown in Ref.~\cite{Du:2017zvv}, the amplitudes with
the two  $D_0^*$ states are fully consistent with the LHCb measurements of the reaction $B^-\to
D^+\pi^-\pi^-$, which are at present the best data providing access to the $D\pi$ system and thus to
the nonstrange scalar charm mesons. This information is encoded in the so-called angular momenta, which
are discussed in detail in the LHCb paper~\cite{Aaij:2016fma}. 

The theoretical framework to analyse this process is based on the unitarized
chiral effective Lagrangian, Eq.~(\ref{LBDpipi}), where one pion is fast and the other
participates in the $D\pi$ final-state interactions (for more details, see Ref.~\cite{Du:2017zvv}).
To be specific, consider the reaction $B^-\to D^+ \pi^-\pi^-$. For sufficiently low energies in the $D\pi$
system, it suffices to include the lowest partial waves (S,P,D), so we can write the decay amplitude as
\begin{equation}
 {\cal{A}}(B^- \to D^+\pi^-\pi^-) = \sum_{L=0}^2 \sqrt{2L+1} {\cal{A}}_L(s) P_L(z)  \, ,
\end{equation}
where ${\cal{A}}_{0,1,2}(s)$ correspond to the amplitudes with $D^+\pi^-$ in the S- , P-  and
D-waves, respectively, and  $P_L(z)$ are the Legendre polynomials. For the P- and D-wave amplitudes
we use the  same Breit-Wigner form as in the LHCb analysis~\cite{Aaij:2016fma}, containing the $D^*$ and $D^*(2680)$
mesons in the P-wave and the $D_2(2460)$ in the D-wave. For the S-wave, however, we employ
\begin{eqnarray}
  {\cal{A}}_{0}(s) &=& A \bigg\{ E_\pi \!\left[ 2+
  G_{1}(s)\left(\frac53 T^{1/2}_{11}(s) + \frac{1}{3}
  T^{3/2}(s) \right) \right] \nonumber\\ 
  &&+ \frac{1}{3} E_{\eta} G_{2}(s) T^{1/2}_{21}(s)
  + \sqrt{\frac23 } E_{\bar K} G_{3}(s) T^{1/2}_{31}(s) \bigg\}
  + B E_\eta G_{2}(s) T^{1/2}_{21}~,
\label{eq:amp}
\end{eqnarray}
where  $A$ and $B$ are two independent couplings following from SU(3) flavor symmetry (i.e. combinations
of the LECs $c_i$,  $A=\sqrt{2}(c_1+c_4)/F_\phi$ and $B=2\sqrt{2}(c_2+c_6)/(3F_\phi)$, with $F_\phi$ the GB decay constant),
and $E_{\pi,\eta,\bar K}$ are
the energies of the light mesons. Further, the $T^{I}_{ij}(s)$ are the S-wave scattering amplitudes for the
coupled-channel system with total isospin $I$, where $i,j$ are channel indices with $1,2$ and $3$ referring
to $D\pi$, $D\eta$  and $D_s\bar K$, respectively, and the $G_i(s)$ are the corresponding 2-point loop functions.  These scattering amplitudes are again taken from
Ref.~\cite{Liu:2012zya} where also all the other parameters were fixed. To filter out the S-wave, the
following (combinations of) angular moments are used:
\begin{eqnarray}
\langle P_0\rangle &\propto& |{\mathcal A}_0|^2 +  |{\mathcal A}_1|^2 +  |{\mathcal A}_2|^2~, \nonumber\\
\langle P_2\rangle &\propto& \frac{2}{5}|{\mathcal A}_1|^2 +  \frac{2}{7}|{\mathcal A}_2|^2
                          +\frac{2}{\sqrt{5}}|{\mathcal A}_0||{\mathcal A}_2| \cos(\delta_2-\delta_0)~,\nonumber\\
\langle P_{13}\rangle &=& \langle P_1\rangle -  \frac{14}{9} \langle P_3\rangle
               \propto  \frac{2}{\sqrt{3}}|{\mathcal A}_0||{\mathcal A}_1| \cos(\delta_1-\delta_0)~,
\end{eqnarray}
with $\delta_0, \delta_1, \delta_2$ the S-, P-, D-wave phase shift, respectively.
\begin{figure}[t!]
  \begin{center}
   \includegraphics[width=0.31\linewidth]{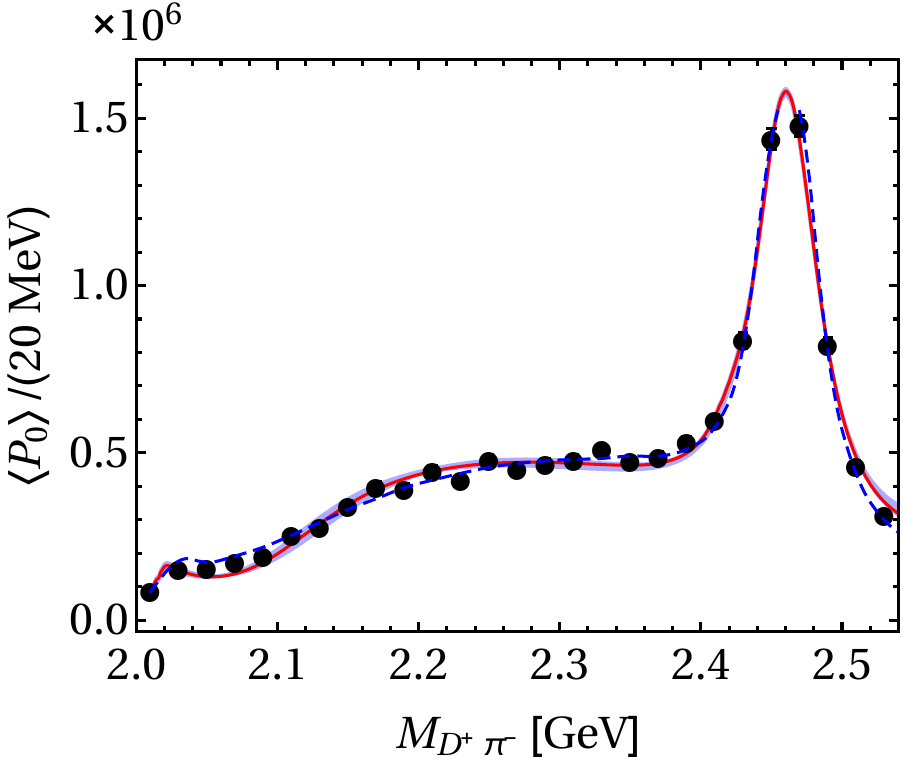} \hfill
  \includegraphics[width=0.32\linewidth]{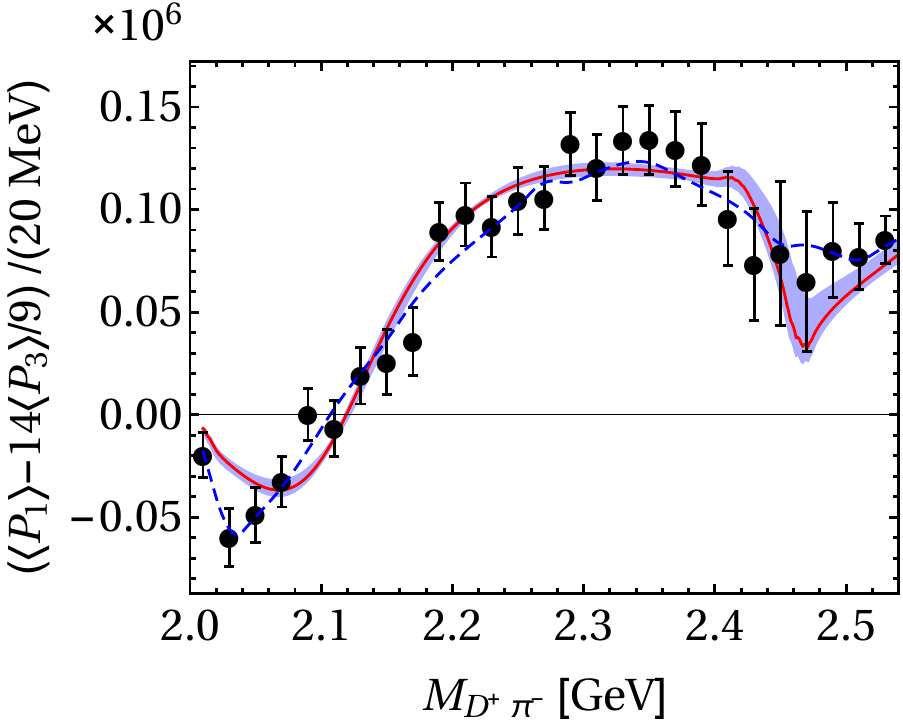} \hfill
  \includegraphics[width=0.31\linewidth]{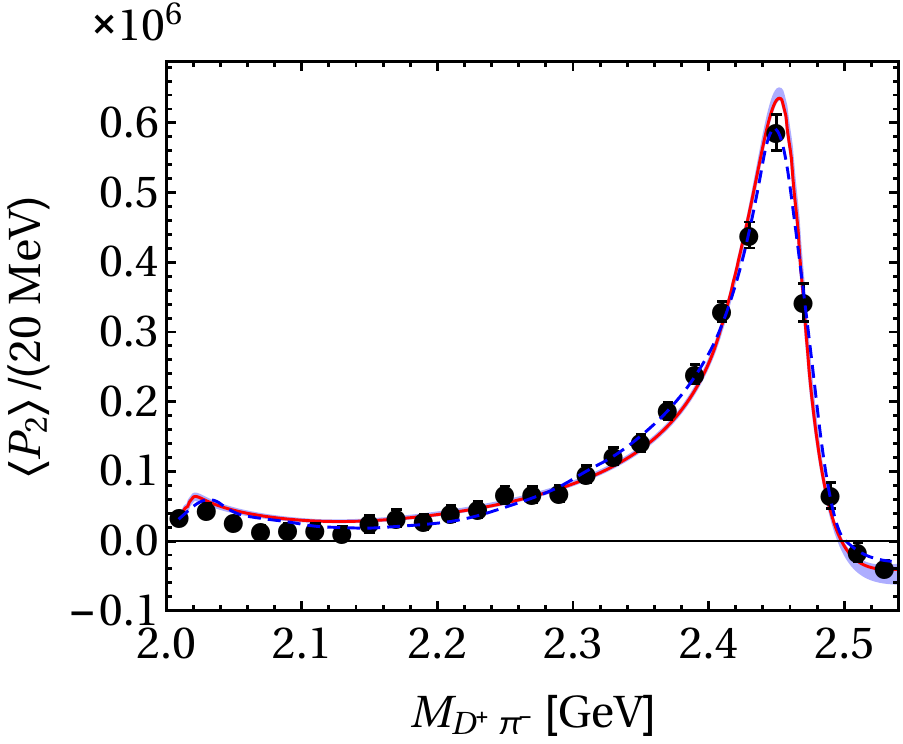}
 \end{center}
  \caption{
    Fit to the LHCb data for the angular moments $\langle P_0\rangle$, $\langle P_{13}\rangle$ and
    $\langle P_2\rangle$ for the  $B^-\to D^+\pi^-\pi^-$ reaction~\cite{Aaij:2016fma}. 
    The largest error among $\langle P_1\rangle$ and $14\langle P_3\rangle/9$ in each bin is taken as
    the error of $\langle P_1\rangle - 14\langle P_3\rangle/9$. The solid lines show  
    the results of~\cite{Du:2017zvv}, with error bands corresponding to the one-sigma uncertainties propagated 
    from the input scattering amplitudes, while the dashed lines stand for the LHCb fit using
    cubic splines for the S-wave~\cite{Aaij:2016fma}.
\label{fig:dpi}}
\end{figure}
%
The  best fit to the LHCb data is shown in Fig.~\ref{fig:dpi} together with their best fit provided  by LHCb
based on cubic splines (dashed lines). The bands in Fig.~\ref{fig:dpi}
reflect the one-sigma errors of the parameters in the scattering amplitudes  
determined in Ref.~\cite{Liu:2012zya}.
It is worthwhile to notice that in $\langle P_{13}\rangle$, where the
$D_2(2460)$ does not play any role, the data show a significant variation
between 2.4 and 2.5~GeV. Theoretically this feature can now be understood as a
signal for  the opening of the $D^0\eta$ and $D_s^+ K^-$ thresholds at 2.413 and 2.462~GeV, respectively,
which leads to two cusps in the amplitude.  This effect is amplified by the higher pole which is relatively
close to the $D_s\bar K$ threshold on the unphysical sheet.
There is some discrepancy between the
chiral amplitude and the data for $\langle P_{13}\rangle$ at low energies: Does this point at a deficit of
the former? Fortunately the LHCb Collaboration provided more detailed information on their S-wave
amplitude in Ref.~\cite{Aaij:2016fma}: In the analysis of the data a series of anchor points
were defined where the strength and the phase of the S-wave amplitude were extracted
from the data. Then cubic splines were used to interpolate between these anchor points.
\begin{wrapfigure}{r}{0.45\textwidth}
\centering
\vspace*{-.1cm}
\includegraphics[width=0.44\textwidth]{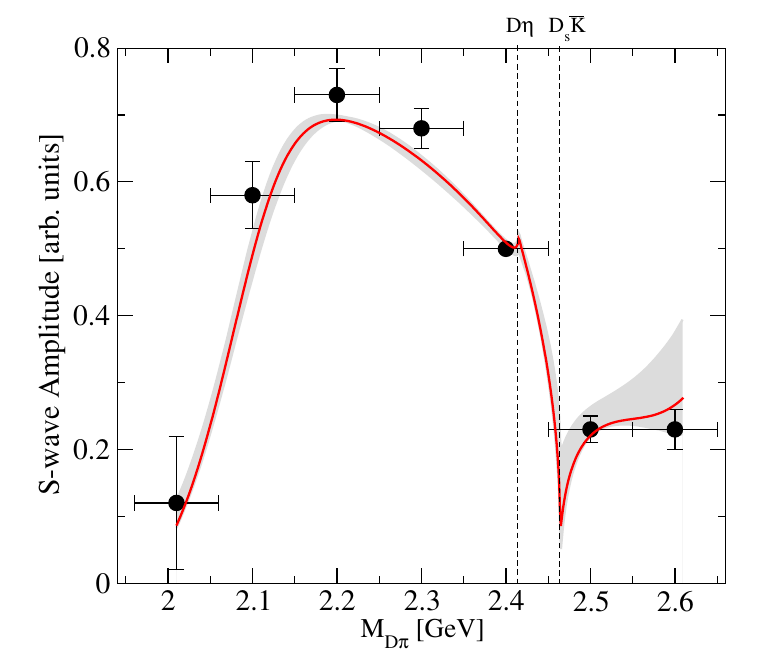}
\caption{Comparison of the S-wave amplitude based on UCHPT to the S-wave anchor points found in the 
  experimental analysis, shown as the data points~\cite{Aaij:2016fma}. The red line gives the best
  fit results and the grey band quantifies the uncertainties that emerged from the fitting procedure.
  The fitting range extends up to 2.55~GeV. The dashed perpendicular lines indicate the location of
  the $D\eta$ and $D_s \bar K$ threshold, respectively.
  \label{fig:dpiSwave}}
\vspace{-2mm}
\end{wrapfigure}
In  Fig.~\ref{fig:dpiSwave} the S-amplitude fixed as described above is compared to the
LHCb anchor points. Not only shows this figure very clearly that the strength of the S-wave amplitude
largely determined by the fits to lattice data is fully consistent with the one extracted from the data
for $B^-\to D^+\pi^-\pi^-$,  the shown amplitude also shows the importance of the $D\eta$ and $D_s \bar K$ cusps
and thus also of the role of the higher pole in the $I=1/2$ and $S=0$ channel even more clearly
than the angular moments discussed above. This clearly highlights  the importance of  a coupled-channel
treatment for this reaction. An updated analysis of the LHC Run-2 data is called for to confirm the
prominence of the two cusps.

LHCb presented also data on $B_s^0\to \bar{D}^0 K^- \pi^+$, which are, however, less precise than the ones
just discussed. Using the same formalism as before, with one different combination of the LECs $c_i$
and the same resonances in the P- and D-wave as LHCb, these data can be well described by  a one parameter fit,
see Ref.~\cite{Du:2017zvv} for more details.  A combined analysis including also data for
$B^0\to \bar{D}^0\pi^-\pi^+$, $B^-\to D^+\pi^-K^-$ and $B_s^0\to \bar{D}^0K^-\pi^+$ performed in
Ref.~\cite{Du:2019oki} gives further credit to this picture.

\subsection{The $K_1$ meson}

Another state that offered support to the two-pole scenario even before the heavy-light mesons just discussed
is the axial-vector meson $K_1(1270)$, which in the quark model is a kaonic excitation with angular momentum one,
$I(J^P) =\frac{1}{2}(1^+)$.
The two-pole nature of the  $K_1(1270)$ was first noted in the study of the scattering of vector mesons
off the Goldstone bosons in a chiral unitary approach at tree level~\cite{Roca:2005nm}. This was further
sharpened in Ref.~\cite{Geng:2006yb}. There,  the high-statistics data from the  WA3 experiment on
$K^- p \to K^-\pi^+\pi^- p$, analyzed by the ACCMOR collaboration~\cite{Daum:1981hb}, were reanalyzed
and shown to favor a two-pole interpretation of the $K_1(1270)$. The authors of Ref.~\cite{Geng:2006yb}
also reanalyzed the traditional K-matrix interpretation of the WA3 data and found that the good fit of
the data obtained there was due to large cancellations of terms of unclear physical interpretation. It
was recently shown how this two-pole scenario can show up in $D$-meson decays, in particular $D^0\to \pi^+ V P$
and $D^+\to \nu e^+ V P$, where $P$ and $V$ are pseudoscalar and vector mesons, respectively~\cite{Wang:2019mph,Wang:2020pyy}.

\section{Discussion and Outlook}
\label{sec:summ}

Let me summarize briefly:
\begin{itemize}
\item The story with the two-pole structure started with the $\Lambda(1405)$, which can now be considered
  as established. Still, the position of lighter pole  close to the $\pi\Sigma$ threshold needs to be
  determined better whereas the higher pole close to the $K^-p$ threshold is pretty well pinned down. 
  It is comforting to note that the re-analysis of the J\"ulich $\bar{K}N$ meson-exchange model from
  the 1990s also confirmed
  the two-pole structure of the $\Lambda(1405)$, see Ref.~\cite{Haidenbauer:2010ch} (and references therein).
  I again point out that approaches, that do not allow for the dynamical generation of resonances, like e.g. the BnGa 
  model, are  insufficient for describing the whole hadron spectrum.
\item Further support of the two-pole scenario comes from charmed baryons. Recently, an analysis of the LHCb 
    data on  $\Lambda_b\to pD^0\pi^-$ in the near-$pD^0$-threshold region   
   also revealed a two-pole structure of the $\Sigma_c(2800)^+$ when isospin-breaking is taken into account
   \cite{Sakai:2020psu}.
\item The spectrum of excited charmed mesons, made from a heavy $c$ quark and a light $u,d,s$
  quark, offers further support of the two-pole structure and the dynamical generation of hadron resonances.
  Here, a beautiful interplay of experimental results, unitarized chiral perturbation theory and lattice QCD
  gives very strong indications that this picture is indeed correct. Further lattice calculations and the measurement
  of the corresponding $B$-mesons will serve as further tests.  
\item  
  This leads to a new paradigm in hadron physics: The hadron spectrum must not be viewed
as  a collection of quark model states, but rather as a manifestation 
of a more complex dynamics that leads to an intricate pattern of various types 
of states that can only be understood by an interplay of theory and experiment
(cf. the light scalar mesons or the states discussed here).
\item
 The dynamical generation of hadron states through hadron-hadron interactions ties together
 nuclear and particle physics, as these molecular compounds bear resemblance to the light
 nuclei, the deuteron, the triton and so on. Therefore, such molecular states were called ``deusons'' 
 by T\"ornquist, one of the pioneers in the field of hadronic molecules~\cite{Tornqvist:1993ng}.
 \end{itemize}
 
From all this, it is rather obvious that the PDG tables published in the RPP must undergo
a drastic change and finally acknowledge the two-pole structure in the main listings, not just
in the review section. Time will tell how long this necessary  change will take.

\vspace{6pt} 

\section*{Acknowledgements}

I am grateful to Feng-Kun Guo, Maxim Mai and Jos\'e Oller for comments on the manuscript and
FKG and MM for providing me with some of the figures shown. I also thank all my collaborators on the issues discussed
here for sharing their insights. Finally, I am grateful to Dubravko Klabucar for giving me the opportunity to
write up these thoughts. This research was funded in part by Deutsche Forschungsgemeinschaft (TRR 110,
``Symmetries and the Emergence of Structure in QCD''), the Chinese Academy of Sciences
 (CAS) President's International Fellowship Initiative (PIFI) (grant no. 2018DM0034) and
 VolkswagenStiftung (grant no. 93562).





\end{document}